\documentclass[12pt%,
]{iopart}

\usepackage{graphicx}
\usepackage{siunitx}
\usepackage{color}
\usepackage{tabularx}
\begin{document}

\title{Recombination dynamics of type-II excitons in (Ga,In)As/GaAs/Ga(As,Sb) heterostructures}

\author{S~Gies, B~Holz, C~Fuchs, W~Stolz and W~Heimbrodt}
\ead{wolfram.heimbrodt@physik.uni-marburg.de}
\address{Department of Physics and Material Sciences Center, Philipps-Universit\"{a}t Marburg, Renthof 5, 35032 Marburg, Germany}

\date{\today}

%%%%%%%%%%%%%%%%%%%%%%%%%%%%%%%%%%%%%%%%%%%%%%%%%%%%%%%%%%%%%%%%%%%%%%%%%%%%%%%%%%%%%%%%%%%%%%%%%%%%%%%%%%%%%%%%%%%%%%%%%%%%%%%%%%%%%%%%%%%%%%%%%%%%%
\begin{abstract}
(Ga,In)As/GaAs/Ga(As,Sb) multi-quantum well heterostructures have been investigated using continuous wave and time-resolved photoluminescence spectroscopy at various temperatures.
A complex interplay was observed between the excitonic type-II transitions with electrons in the (Ga,In)As well and holes in the Ga(As,Sb) well and the type-I excitons in the (Ga,In)As and Ga(As,Sb) wells. 
The type-II luminescence exhibits a strongly non-exponential temporal behavior below a critical temperature of $T_c=\SI{70}{K}$.
The transients were analyzed in the framework of a rate-equation model. It was found that the exciton relaxation and hopping in the localized states of the disordered ternary Ga(As,Sb) are the decisive processes to describe the dynamics of the type-II excitons correctly. 

\end{abstract}
\maketitle

%%%%%%%%%%%%%%%%%%%%%%%%%%%%%%%%%%%%%%%%%%%%%%%%%%%%%%%%%%%%%%%%%%%%%%%%%%%%%%%%%%%%%%%%%%%%%%%%%%%%%%%%%%%%%%%%%%%%%%%%%%%%%%%%%%%%%%%%%%%%%%%%%%%%%
\section{Introduction}
The (Ga,In)As/Ga(As,Sb) material system is used for a wide variety of applications nowadays.
For example (Ga,In)As/Ga(As,Sb) structures can be used as active medium for mid-infrared lasing.\cite{Pan2010,Chang2014,Pan2013,Huang2009}
Furthermore, applications as light sources in the $\SI{1.6}{\micro m}$ based on (Ga,In)As quantum dots capped with Ga(As,Sb) have been realized.\cite{Ripalda2005}
Very recently, the (Ga,In)As/Ga(As,Sb) material system has been used to make vertical-external-cavity surface emitting lasers emitting light at 
$\SI{1.2}{\micro m}$ using the type-II band alignment.\cite{Berger2015,Gies2015,Moeller2016} 
To further improve these devices profound knowledge about the basic properties and processes happening in these materials are needed.
Especially the recombination dynamics of this type-II system is an important issue and need to be studied carefully. Few reports exist to describe the recombination processes.\cite{Tatebayashi2008,Morozov2014}
In this work we aim to give a thorough analysis of the recombination processes in (Ga,In)As/GaAs/Ga(As,Sb) heterostructures.
We investigate not only the decay behavior experimentally by means of time-resolved photoluminescence, but provide also a rate-equation model to reveal the important underlying processes of recombination, relaxation and tunneling. Furthermore, this comprehensive study addresses the changes in the type-II luminescence and recombination kinetics at different temperatures, detection energies and discusses the influence of an internal barrier of GaAs. The exciton relaxation turned out to be important in those type II structures. The barrier width can be used to selectively change tunneling and recombination times while keeping the relaxation process constant. This way we were able to determine independently all the important parameters of the type II exciton dynamics.

%%%%%%%%%%%%%%%%%%%%%%%%%%%%%%%%%%%%%%%%%%%%%%%%%%%%%%%%%%%%%%%%%%%%%%%%%%%%%%%%%%%%%%%%%%%%%%%%%%%%%%%%%%%%%%%%%%%%%%%%%%%%%%%%%%%%%%%%%%%%%%%%%%%%%
\section{Experimental}
Our samples were grown on exact GaAs (001) substrates using metal-organic vapor-phase epitaxy (MOVPE).
The sample growth was carried out in an AIXTRON AIX 200 GFR (Gas Foil Rotation) reactor system at a pressure of \SI{50}{mbar} and using H$_2$ as carrier gas.
The native oxide layer was removed from the substrates prior to the sample growth by applying a tertiarybutylarsine (TBAs)-stabilized bake-out procedure.
The following growth of the active region was carried out at a temperature of \SI{550}{\celsius} using triethylgallium (TEGa) and trimethylindium (TMIn) as group-III and TBAs and triethylantimony (TESb) as group-V precursors.
The active region consists of a \mbox{5 $\times$ multiple double quantum well heterostructure}.
Each repetition is composed of a \SI{5.2}{nm} thick (Ga$_x$,In$_{1-x}$)As layer which is followed by a GaAs interlayer of variable thickness d.
The active region is completed by a second \SI{5.0}{nm} thick quantum well consisting of Ga(As$_{1-y}$,Sb$_y$) which is followed by a \SI{50}{nm} thick GaAs barrier.

\begin{table}%
\caption{Compositions and interlayer thicknesses of the investigated samples.}
\begin{center}
\begin{tabular}{ccc}
\hline
\hline
d (nm)& x$_{In}$ (\%) & y$_{Sb}$ (\%) \\
\hline
0.4 & 20.7 & 21.1 \\
1.5 & 21.5 & 21.7 \\
3.5 & 21.0 & 23.8 \\
4.8 & 21.0 & 23.3 \\
\hline
\hline
\end{tabular}
\end{center}
\label{tab:tab1}
\end{table}

The cw-photoluminescence (PL) spectra have been measured using a liquid nitrogen cooled Ge-detector and a \SI{0.5}{m} grating spectrometer.
A frequency-doubled solid state laser at \SI{532}{nm} provided the light for the excitation of the sample.
The time resolved PL measurements were performed using a frequency-doubled Nd:YAG at \SI{532}{nm} and a repetition rate of \SI{10}{Hz}.
The PL was detected using a \SI{0.25}{m} grating spectrometer and a thermoelectrically cooled InP/(In,Ga)(As,P) photomultiplier.
Due to the temporal linewidth of the laser the time-resolution of our setup is \SI{4}{ns}.

%%%%%%%%%%%%%%%%%%%%%%%%%%%%%%%%%%%%%%%%%%%%%%%%%%%%%%%%%%%%%%%%%%%%%%%%%%%%%%%%%%%%%%%%%%%%%%%%%%%%%%%%%%%%%%%%%%%%%%%%%%%%%%%%%%%%%%%%%%%%%%%%%%%%%
\section{Results and Discussion}
\begin{figure}[!ht]
  \includegraphics[width=8.5cm]{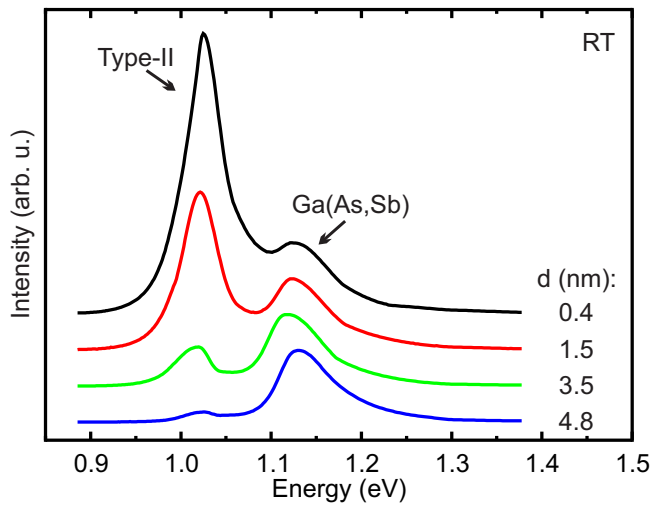}
  \caption{Room-temperature PL spectra for the samples with different interlayer thickness d.
	The spectra are normalized to the Ga(As,Sb) emission and shifted vertically for clarity.}
  \label{fig:fig1}
\end{figure}

The room-temperature PL spectra are depicted in figure \ref{fig:fig1} for the four samples with different interlayer thicknesses.
The spectra are normalized to the respective Ga(As,Sb) type-I emission around \SI{1.15}{eV} \cite{Antypas1970,Nahory1977} and shifted vertically for clarity. 
At approx. \SI{1.025}{eV} an additional peak can be seen in all spectra.
This is due to the recombination of electrons in the (Ga,In)As and holes in the Ga(As,Sb).
The intensity of this type-II emission increases with respect to the type-I PL with decreasing interlayer barrier thickness d.
Such a behavior can easily be understood, because with decreasing separation of the electrons and holes the overlap of their wavefunctions is increased and therefore the recombination probability. The slight deviation in peak position between the different samples is explained by small variations in layer composition (c.f. table \ref{tab:tab1}). 

\begin{figure}[h!t]
\includegraphics[width=8.5cm]{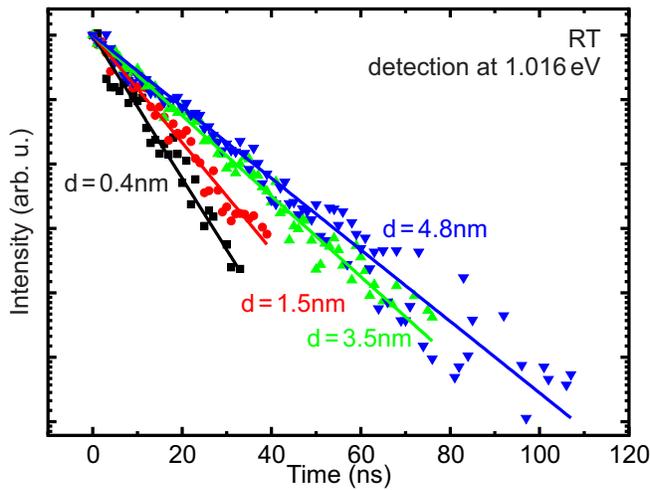}
\caption{Room-temperature decay curves of the type-II PL for the samples with differently thick interlayers measured at the PL maximum at $E_{Det} = \SI{1.016}{eV}$.}
\label{fig:fig2}
\end{figure}

To further analyze the behavior of the type-II PL its transients are presented in figure \ref{fig:fig2}.
These were measured at the maximum of the type-II emission and are normalized to unity.
Because of the spectrally close Ga(As,Sb) PL the depicted transients are taken \SI{7}{ns} after the excitation laser pulse had its maximum.
This guarantees that only the type-II PL is analyzed, because the GaAsSb emission has decayed to a negligible level as its radiative lifetime is in the picosecond range.
The transients were measured at a detection energy of $E_{Det} = \SI{1.016}{eV}$, which corresponds to the maximum of the type-II PL.

\begin{figure}[h!t]
\includegraphics[width=8.5cm]{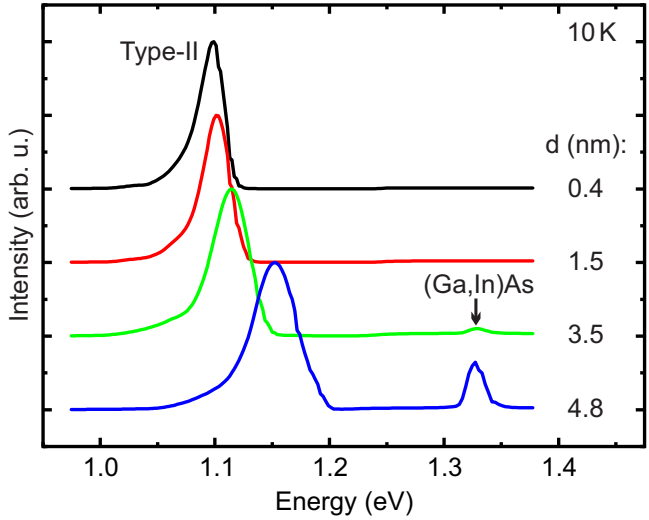}
\caption{Low-temperature spectra of the (Ga,In)As/GaAs/Ga(As,Sb) heterostructures.
The spectra are normalized to the type-II emission and shifted vertically.}
\label{fig:fig3}
\end{figure}

At room-temperature the decay of the type-II PL exhibits a monoexponential behavior with lifetimes in the ns-range, which is typical for type-II transitions.
The PL lifetime of a given transient in figure \ref{fig:fig2} can easily be determined. 
For the thinnest internal barrier of $d = \SI{0.4}{nm}$ we find an $e^{-1}$-time of $\tau_{\SI{0.4}{nm}} = \SI{9}{ns}$.
The radiative lifetime increases with increasing barrier thickness to values of $\tau_{\SI{1.5}{nm}} = \SI{12}{ns}$, $\tau_{\SI{3.5}{nm}} = \SI{16}{ns}$, and $\tau_{\SI{4.8}{nm}} = \SI{18}{ns}$ for the thickest barrier.
Considering the scatter of the data-points in figure \ref{fig:fig2} the uncertainty for all these lifetimes is $\pm \SI{1}{ns}$. 
This increase in decay time is related to the decrease of type-II PL intensity (c.f. figure \ref{fig:fig1}).
Due to the reduced overlap of electron and hole wave-functions the radiative transition probability decreases and the radiative lifetime increases, respectively.
To further analyze the behavior of the type-II PL we present the \SI{10}{K} spectra of the four samples in figure \ref{fig:fig3}.

\begin{figure}
\includegraphics[width=8.5cm]{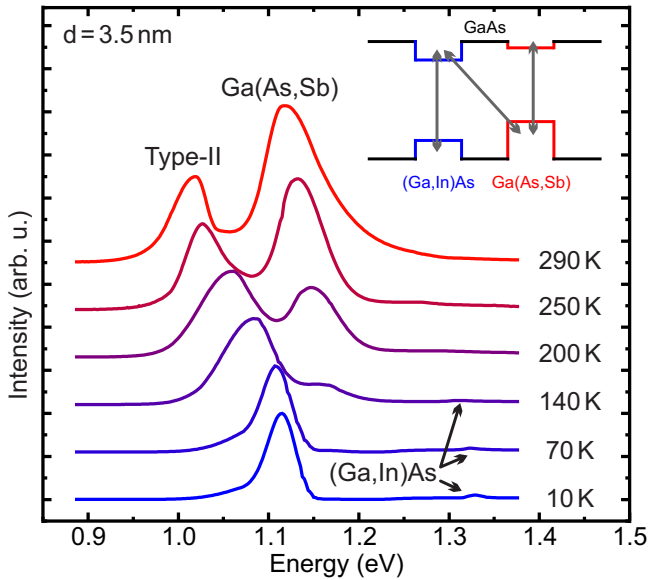}
\caption{Temperature-dependent PL spectra of the sample with $d = \SI{3.5}{nm}$.
The spectra are normalized to the type-II peak and shifted vertically for clarity.
In the upper right corner depicted are the band edges of our heterostructures with the growth direction being from left to right.}
\label{fig:fig4}
\end{figure}

The PL spectra at low temperature are normalized again to the type-II emission. 
Besides the type-II PL, we detect an emission band at \SI{1.33}{eV} for the samples with thickest internal barriers.
This emission band is characteristic for (Ga,In)As.\cite{Goetz1983}
Surprisingly, this emission only occurs at low temperatures and is not observable at room-temperature.
Furthermore, the (Ga,In)As emission is rather weak and only observable for the thickest interlayer barriers. 
We will come back to this behavior later. 
Additionally, the Ga(As,Sb) emission that was clearly visible at RT cannot be seen in the \SI{10}{K} spectra.
To understand this behavior, we exemplary investigate the temperature-dependence of the PL of the sample with $d = \SI{3.5}{nm}$.
The spectra are given in figure \ref{fig:fig4}.

From the temperature-dependent spectra one can directly see, that the Ga(As,Sb) PL signal vanishes below temperatures of \SI{140}{K}.
Furthermore, the (Ga,In)As emission appears below this temperature. To understand this behavior it is helpful to take a look at the band structure of the (Ga,In)As/GaAs/Ga(As,Sb) heterostructure.
It is depicted in the inset of figure \ref{fig:fig4}. The growth direction is from left to right.
Both, the (Ga,In)As\cite{Zubkov2004} and the Ga(As,Sb)\cite{Morozov2014}\ have a type-I band alignment with respect to GaAs.
The band-discontinuity between (Ga,In)As and Ga(As,Sb), however, is of type-II.\cite{Morozov2014,Hu1998}
After relaxation, the energetically most favorable states for the electrons are in the conduction band (CB) of the (Ga,In)As well and for the holes the Ga(As,Sb) well. 
From this picture one would expect to see a dominant type-II luminescence between electrons in the (Ga,In)As and the holes in the Ga(As,Sb).
Indeed, this transition dominates the spectrum at low temperatures (c.f. figure \ref{fig:fig3}).
Why do we see (Ga,In)As type-I luminescence at low temperatures? To answer this question, we need to consider the following.
The excitation of our samples takes place at an energy of \SI{2.33}{eV}, which is well above the bandgap of the GaAs barriers in our heterostructure.
Therefore, we create a lot of charge-carriers that can relax into both QWs. The holes in the (Ga,In)As then have two possibilities. They may recombine radiatively with the electrons in the (Ga,In)As, which yields the PL at \SI{1.33}{eV}, or they can tunnel into the energetically more favorable states in the Ga(As,Sb). 
At low temperatures the phonon assisted tunneling probability is reduced, which allows for an observation of the (Ga,In)As line. The fact that the (Ga,In)As PL increases with increasing interlayer thickness (c.f. figure \ref{fig:fig3}) strengthens this argument as the tunneling probability is lower for a thicker barrier.
The absence of the Ga(As,Sb) emission in the \SI{10}{K} spectrum is in accordance with this model, since the electron-tunneling is much faster than hole-tunneling because the effective mass of the electrons is smaller but even the tunneling barrier is lower in our structure. Obviously, the Ga(As,Sb) luminescence cannot be observed at \SI{10}{K} because the electrons tunnel on a timescale faster than the exciton recombination time in Ga(As,Sb). 
Increasing the temperature to \SI{70}{K} leads to a drop in the PL intensity of the (Ga,In)As. This is now  due to the very effective phonon assistance in the hole-tunneling.
Above \SI{140}{K} the (Ga,In)As emission vanishes completely, because of the lack of holes in the QW.

Surprisingly, the Ga(As,Sb) emission starts to occur in the spectrum even though the electrons should still tunnel to the (Ga,In)As QW very fast.
This is due to the fact that electrons can occupy higher electronic states by thermal excitation and can return this way to the Ga(As,Sb) QW at higher temperatures. 
The same process is not possible for the holes.
Eventually, the type-I PL in the Ga(As,Sb) is strongest at room temperature, although the electrons are spread over both wells and the holes are still dominantly in the Ga(As,Sb) well. The transition probability of the type-I transition is of course much bigger, due to the strong overlap of electron and hole wavefunctions, compared to the type-II transition of the charge transfer (CT) excitons.

\begin{figure}
\includegraphics[width=8.5cm]{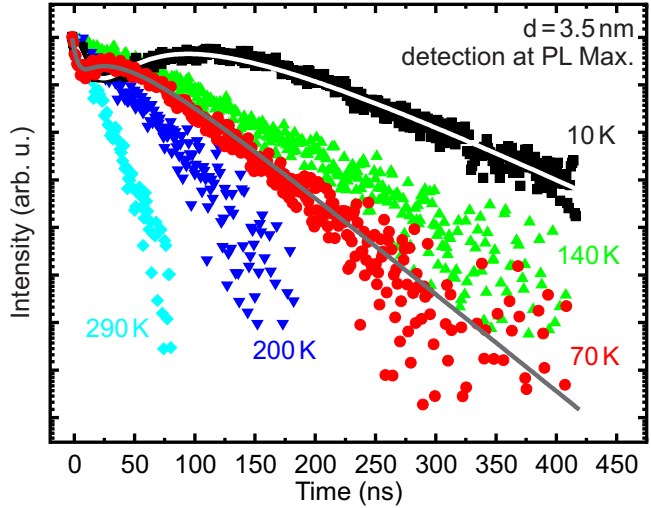}
\caption{Transients of the type-II PL at different temperatures for the sample with $d = \SI{3.5}{nm}$.
The curves are normalized to unity and were taken at the maximum of the type-II peak.
The solid curve are the fitted transients obtained by our tri-exponential model (see below for details).}
\label{fig:fig5}
\end{figure}

This pronounced behavior of the charge carriers as a function of ambient temperature should also strongly influence the recombination dynamics of the excitons. We therefore performed time resolved studies of the type-II recombination in the temperature range between \SI{10}{K} and \SI{290}{K}.
The decay curves for the sample with $d = \SI{3.5}{nm}$ are shown in figure \ref{fig:fig5}.
These transients were taken at the maximum of the type-II PL. 
For temperatures between \SI{290}{K} and \SI{140}{K} the decay is still mono-exponential. 
Starting at $\tau_{290\,K} = (16 \pm 1)\,ns$ the decay time increases, however, to $\tau_{200\,K} = (37 \pm 5)\,ns$ and $\tau_{140\,K} = (88 \pm 5)\,ns$ for \SI{140}{K}.
The increasing decay time with decreasing temperature is typical for exciton recombination. This is explained by reduced electron-phonon-coupling and reduced nonradiative losses.
These explanation should very much apply to our type-II recombination process.  
Interestingly, below \SI{140}{K} the shape of the transients becomes distinctly different.
The mono-exponential decay behavior disappears and a delayed increase can be seen.
The time dependence consists of three parts now.
In the beginning there is a rather fast decay of the type-II PL.
This is followed by a rise of the curve.
The transient reaches a maximum and declines later with a slow decay time.

\begin{figure}
\includegraphics[width=8.5cm]{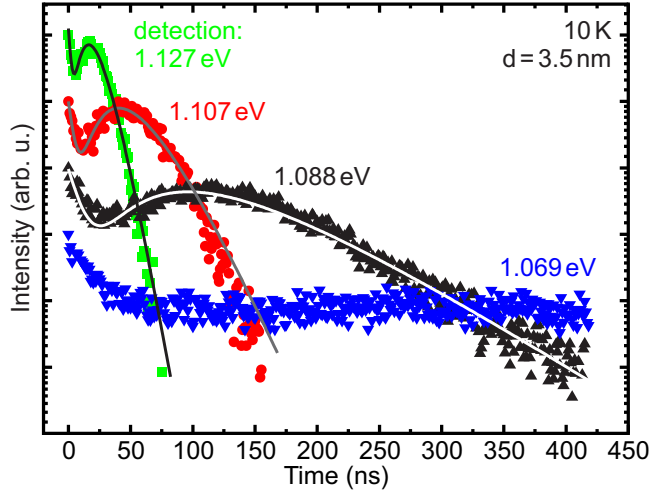}
\caption{Normalized decay curves of the type-II PL for the samples with $d = \SI{3.5}{nm}$ at a temperature of \SI{10}{K}.
The detection was shifted across the spectrum and the transients are shifted vertically for clarity.
The black decay curve is the transient previously discussed in figures \ref{fig:fig5}.
The solid lines are fitted to the respective curve using our tri-exponential model.}
\label{fig:fig7}
\end{figure}

A similar behavior was found by Morozov et al. investigating (Ga,In)As/Ga(As,Sb) QWs at low temperatures.\cite{Morozov2014} 
The respective lifetimes they measured were considerably shorter than ours, which might be explained by the absence of an internal barrier in their samples.
The threepart decay curve was explained by considering the screening effect of the charge carriers in higher type-I states that reduces the band bending due to the type-II excitons.
In the first time regime this screening is reduced as the type-I excitons recombine and the point where the transient reaches its local minimum is reached after the typical type-I decay time.
The following incline in the transient is then caused by the reduction of the band bending due to the recombination of the type-II excitons and the resulting red-shift of the emission line.
Finally, the the last part of the decay curve represents the type-II decay time.
This model \cite{Morozov2014}, however, cannot explain the transients observed here.
This becomes particularly obvious by considering transients at different PL energies. The decay curves detected at different positions across our PL spectrum. The respective decay curves are depicted in figure \ref{fig:fig7}.

It can be seen, that both the minimum and the maximum is shifted to later times the lower the detection energy was. The only difference between the curves is an energy relaxation process.  
It seems to be necessary to develop a model including relaxation processes.
The relaxation itself comprises various steps.
The excitation energy is above the GaAs bandgap. Electron and hole relaxation can take place.
Furthermore, the QWs are made of ternary materials yielding a certain amount of microscopic alloy disorder.
The disorder results in a potential landscape of localized states in the QWs.
Such localized states result in a hopping mobility of the excitons, which is particularly important at low temperatures.
We have shown in an earlier paper, that hopping relaxation in disordered systems can easily reach hundreds of ns.\cite{Niebling2008}   
To adequately describe the decay behavior at \SI{70}{K} and below, we develop in the following a kinetic model taking three processes into account.
First, the excitons ($n_{In}$) in the (Ga,In)As can either recombine radiatively in their well ($\tau_{In}$) or form type-II excitons by tunneling of the holes to the Ga(As,Sb) with the probability $w_T=1/\tau_T$. This yields the differential equation \ref{eq:eq1}.

\begin{equation}
\frac{d n_{In}}{dt} = -\frac{n_{In}}{\tau_{In}} -w_T \cdot n_{In}
\label{eq:eq1}
\end{equation}

The hole tunneling feeds higher states ($n_H$) in the Ga(As,Sb) QW. The excitons can then relax into lowest states ($n_i$) with a characteristic time of $\tau_R$.
Such single relaxation time is of course a simplification and just a mean time for the relaxation including hopping processes between deep localized states that have a certain distribution in energy. 

The temporal evolution of the occupation of the higher type-II exciton states is described by equation \ref{eq:eq2}.

\begin{equation}
\frac{d n_H}{dt} = w_T \cdot n_{In} -\frac{n_H}{\tau_R}
\label{eq:eq2}
\end{equation}

Equation \ref{eq:eq3} describes then the temporal evolution of the lowest exciton states that are responsible for the type-II luminescence observed in our experiments.
The corresponding lifetime is $\tau_i$.

\begin{equation}
\frac{d n_i}{dt} = \frac{n_H}{\tau_R} -\frac{n_i}{\tau_i}
\label{eq:eq3}
\end{equation}

The solution of these three coupled differential equations yields a tri-exponential function of the form:

\begin{equation}
n(t) = A \cdot \exp(-\frac{t}{\tau_{eff}}) - B \cdot \exp(-\frac{t}{\tau_R}) + C \cdot \exp(-\frac{t}{\tau_i}).
\label{eq:eq4}
\end{equation}

In equation \ref{eq:eq4} $\tau_{eff}^{-1} = \tau_{In}^{-1} + \tau_T^{-1}$ denotes the effective decay time of excitons in the (Ga,In)As.
The term with $\tau_R$ describes the exciton relaxation and hopping and is responsible for the delayed increase of our PL transients.
As can be seen by the full lines in figure \ref{fig:fig5} a perfect fit is possible to all the experimental transients.   
For the sample with $d = \SI{3.5}{nm}$ at \SI{10}{K} we got a relaxation time $\tau_{R} = \SI{55}{ns}$ and a type-II recombination time $\tau_{i} = \SI{83}{ns}$. 
The times at T=\SI{70}{K} are $\tau_{R} = \SI{27}{ns}$ and $\tau_{i} = \SI{49}{ns}$, respectively. The effective exciton time in (Ga,In)As reaches the experimental time resolution of our setup at \SI{70}{K}. Nevertheless, we found a tendency with increasing temperature from $\tau_{eff}= \SI{20}{ns}$ at \SI{10}{K} to $\tau_{eff} = \SI{4}{ns}$ for \SI{70}{K}. 
Increasing the temperature makes all times faster due to the enhanced electron-phonon-coupling as already mentioned. 

At a glance it seems to be surprising, that the radiative lifetime $\tau_i = \SI{88}{ns}$ at $T = \SI{140}{K}$ is higher than the value for $T = \SI{70}{K}$. 
This can be explained taking into account the potential fluctuation in our strongly disordered system.
At low temperatures the CT-excitons are rather localized with a reduced lifetime the stronger the localization gets.
This behavior has been found for disordered materials in general.\cite{Woscholski2016}
Above the mobility edge the lifetime is obviously longer, but reduces also with increasing temperatures. 

\begin{figure}
\includegraphics[width=8.5cm]{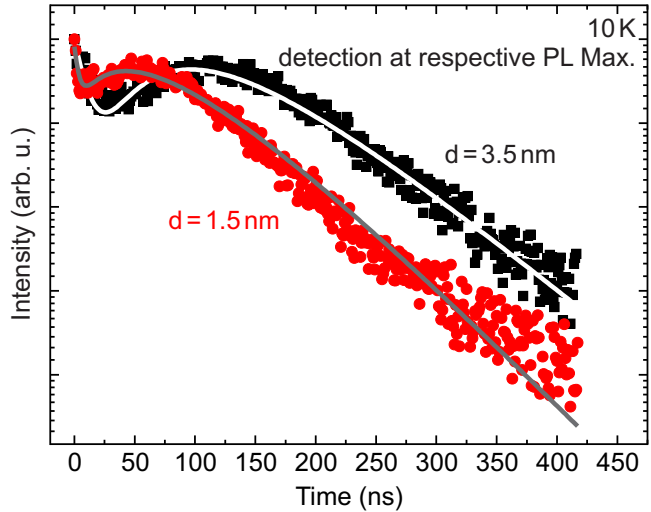}
\caption{Normalized decay curves of the type-II PL for the samples with $d = \SI{1.5}{nm}$ (red) and $d = \SI{3.5}{nm}$ (black).
The detection was done at the respective maximum of the PL at an ambient temperature of \SI{10}{K}.
The fits using our tri-exponential model are given as the solid lines.}
\label{fig:fig6}
\end{figure}

To prove the model, we have performed time-resolved measurements also at the sample with a thinner interlayer of $d = \SI{1.5}{nm}$.
The transients of both samples are depicted in figure \ref{fig:fig6}. One can directly see that both, the local minimum and the maximum occur at later times for the sample with thicker internal barrier.
The black transient is the same curve as in figure \ref{fig:fig5}. 
Our fit yields $\tau_{eff} = \SI{4}{ns}$, $\tau_{R} = \SI{55}{ns}$ and $\tau_{i} = \SI{65}{ns}$ for $d = \SI{1.5}{nm}$.
By decreasing the internal barrier thickness $\tau_{eff}$ decreases substantially, since the hole tunneling gets faster. Even the CT-exciton recombination time $\tau_{i}$ decreases strongly as expected by the increasing dipole matrix element. The relaxation time $\tau_{R}$ remains constant as it is not influenced by changing the barrier thickness. 
This results vigorously support our model and make clear that the CT-exciton PL behavior is indeed strongly influenced by relaxation processes.

Finally, we want to test our model by evaluating the changes in the decay times by moving the detection across the PL spectrum.
The decay curves detected at different positions across our PL spectrum are depicted in figure \ref{fig:fig7}.
The curves are shifted vertically and the measurement was done at \SI{10}{K}.
The black transient measured at a detection energy of $E_{det.} = \SI{1.088}{eV}$ is the same as the black curve in figures \ref{fig:fig5} and \ref{fig:fig6}.
It can be seen that the minimum and the maximum of the transients are shifted to later times as the detection energy is decreased.
The resulting times from the tri-exponential fit are summarized in table \ref{tab:tab2}.
For the transient taken at $E_{det.} = \SI{1.069}{eV}$ no unique fit was possible.

\begin{table}%
\caption{Decay times for the transients given in figure \ref{fig:fig7}.}
\begin{center}
\begin{tabular}{cccc}
\hline
\hline
$E_{det.}$ (eV) & $\tau_{eff}$ (ns) & $\tau_{R}$ (ns) & $\tau_{i}$ (ns) \\
\hline
1.127 & 7.6 & 7.9 & 8 \\
1.107 & 15 & 20 & 21 \\
1.088 & 20 & 55 & 83 \\

\hline
\hline
\end{tabular}
\end{center}
\label{tab:tab2}
\end{table}

It is interesting to note, that the relaxation time and the recombination time increase with decreasing the detection energy (see table\ref{tab:tab2}).
Both changes can be explained in the framework of our model. The higher the detection energy the less relaxation happened and the respective relaxation time is short.
The CT-exciton recombination time is shorter the higher the detection energy, because the relaxation probability is a second loss channel, e.g. the lower the energy the longer the CT exciton lifetime. 
Even $\tau_{eff}$ is slightly shorter the higher the detection energy is. At $E_{det.} = \SI{1.088}{eV}$ we found $\tau_{eff} = \SI{20}{ns}$. This value reduces to $\tau_{eff} = \SI{7.6}{ns}$ for a detection energy of $E_{det.} = \SI{1.127}{eV}$.
This behavior might be caused by a similar relaxation process in the ternary (Ga,In)As.  
Such an relaxation process would be less severe as the (Ga,In)As is less disordered and would lead to shorter times compared to the hopping times induced by the Ga(As,Sb) disorder. 

%%%%%%%%%%%%%%%%%%%%%%%%%%%%%%%%%%%%%%%%%%%%%%%%%%%%%%%%%%%%%%%%%%%%%%%%%%%%%%%%%%%%%%%%%%%%%%%%%%%%%%%%%%%%%%%%%%%%%%%%%%%%%%%%%%%%%%%%%%%%%%%%%%%%%
\section{Conclusion}
In summary, we have investigated the continuous-wave and time-resolved PL of (Ga,In)As/GaAs/Ga(As,Sb) heterostructures with an intermediate barrier of variable thickness between \SI{0.4}{nm} and \SI{4.8}{nm}. At room-temperature we could observe a bright luminescence from the type-II exciton recombination. Additionally, the type-I PL of the Ga(As,Sb) was observed. 
Decreasing the temperature below \SI{140}{K} leads to a disappearance of the Ga(As,Sb) emission. This is because the electrons are no longer thermally excited to the Ga(As,Sb) but are only present in the energetically lowest states of the (Ga,In)As well. At even lower temperatures the (Ga,In)As emission appears, but is very weak for a type-I transition.
The cause for this is the effective hole tunneling from the (Ga,In)As well into the Ga(As,Sb) well, which has a lower tunneling probability than the electron tunneling in opposite direction.
So the (Ga,In)As PL can be observed, while the Ga(As,Sb) peak vanishes. Additionally, the time-resolved measurements reveal an interesting behavior as well.
At high temperatures above \SI{70}{K} the PL decay of the type-II excitons is mono-exponential with radiative lifetimes between $\tau_{290\,K} = (16 \pm 1)\,ns$ and $\tau_{140\,K} = (88 \pm 5)\,ns$.
At \SI{70}{K} and below the shape of the transients changes drastically, since the hopping relaxation between the low lying energy states with relatively low hopping probabilities and respective long times come into play. This behavior has been analyzed in the framework of a rate-equation model.
The temporal evolution of the type-II PL can be explained taking carrier tunneling, relaxation and type-II recombination into account.
We find that at low temperature the relaxation and hopping processes in the Ga(As,Sb) are important and determine the shape of the decay curves.
It is shown that we were able to describe not only the temperature dependence, but also the behavior of samples with differently thick inner barriers as well as the changes of the transients as a function of detection wavelength.

%%%%%%%%%%%%%%%%%%%%%%%%%%%%%%%%%%%%%%%%%%%%%%%%%%%%%%%%%%%%%%%%%%%%%%%%%%%%%%%%%%%%%%%%%%%%%%%%%%%%%%%%%%%%%%%%%%%%%%%%%%%%%%%%%%%%%%%%%%%%%%%%%%%%%
\ack
The work is a project of the Son\-der\-for\-schungs\-be\-reich 1083 funded by the Deutsche Forschungsgemeinschaft (DFG).
S.G. gratefully acknowledges financial support of the DFG in the framework of the GRK 1782.

%%%%%%%%%%%%%%%%%%%%%%%%%%%%%%%%%%%%%%%%%%%%%%%%%%%%%%%%%%%%%%%%%%%%%%%%%%%%%%%%%%%%%%%%%%%%%%%%%%%%%%%%%%%%%%%%%%%%%%%%%%%%%%%%%%%%%%%%%%%%%%%%%%%%%
\section*{References}
\bibliographystyle{iop}
\bibliography{TRPLreferences}

\begin{thebibliography}{10}

\bibitem{Pan2010}
Pan C~H, Lin S~D and Lee C~P 2010 {\it Journal of Applied Physics\/} {\bf 108}

\bibitem{Chang2014}
Chang C~H, Li Z~L, Pan C~H, Lu H~T, Lee C~P and Lin S~D 2014 {\it Journal of
  Applied Physics\/} {\bf 115}

\bibitem{Pan2013}
Pan C~H and Lee C~P 2013 {\it Journal of Applied Physics\/} {\bf 113}

\bibitem{Huang2009}
Huang J~Y~T, Mawst L~J, Kuech T~F, Song X, Babcock S~E, Kim C~S, Vurgaftman I,
  Meyer J~R and Jr A~L~H 2009 {\it Journal of Physics D: Applied Physics\/}
  {\bf 42} 025108

\bibitem{Ripalda2005}
Ripalda J~M, Granados D, González Y, Sánchez A~M, Molina S~I and García J~M
  2005 {\it Applied Physics Letters\/} {\bf 87}

\bibitem{Berger2015}
Berger C, Möller C, Hens P, Fuchs C, Stolz W, Koch S~W, Ruiz~Perez A, Hader J
  and Moloney J~V 2015 {\it AIP Advances\/} {\bf 5}

\bibitem{Gies2015}
Gies S et~al. 2015 {\it Applied Physics Letters\/} {\bf 107}

\bibitem{Moeller2016}
M\"{o}ller C, Fuchs C, Berger C, Ruiz~Perez A, Koch M, Hader J, Moloney J~V,
  Koch S~W and Stolz W 2016 {\it Applied Physics Letters\/} {\bf 108}

\bibitem{Tatebayashi2008}
Tatebayashi J, Liang B~L, Laghumavarapu R~B, Bussian D~A, Htoon H, Klimov V,
  Balakrishnan G, Dawson L~R and Huffaker D~L 2008 {\it Nanotechnology\/} {\bf
  19} 295704

\bibitem{Morozov2014}
Morozov S~V, Kryzhkov D~I, Yablonsky A~N, Aleshkin V~Y, Krasilnik Z~F, Zvonkov
  B~N and Vikhrova O~V 2014 {\it Applied Physics Letters\/} {\bf 104}

\bibitem{Antypas1970}
Antypas G~A and James L~W 1970 {\it Journal of Applied Physics\/} {\bf 41} 2165

\bibitem{Nahory1977}
Nahory R~E, Pollack M~A, DeWinter J~C and Williams K~M 1977 {\it Journal of
  Applied Physics\/} {\bf 48} 1607

\bibitem{Goetz1983}
Goetz K~H, Bimberg D, J\"{u}rgensen H, Selders J, Solomonov A~V, Glinskii G~F
  and Razeghi M 1983 {\it Journal of Applied Physics\/} {\bf 54} 4543

\bibitem{Zubkov2004}
Zubkov V~I, Melnik M~A, Solomonov A~V, Tsvelev E~O, Bugge F, Weyers M and
  Tr\"ankle G Aug 2004 {\it Phys. Rev. B\/} {\bf 70} 075312

\bibitem{Hu1998}
Hu J, Xu X~G, Stotz J~A~H, Watkins S~P, Curzon A~E, Thewalt M~L~W, Matine N and
  Bolognesi C~R 1998 {\it Applied Physics Letters\/} {\bf 73} 2799

\bibitem{Niebling2008}
Niebling T, Rubel O, Heimbrodt W, Stolz W, Baranovskii S~D, Klar P~J and Geisz
  J~F 2008 {\it Journal of Physics: Condensed Matter\/} {\bf 20} 015217

\bibitem{Woscholski2016}
Woscholski R et~al. 2016 {\it Thin Solid Films\/} {\bf 613} 55  spring \{EMRS\}
  symposium: Transport and photonics in group IV-based nano-devices

\end{thebibliography}

%%%%%%%%%%%%%%%%%%%%%%%%%%%%%%%%%%%%%%%%%%%%%%%%%%
\end{document}